# Study of the effect of strain rate on the in-plane shear and transverse compression response of a composite ply using computational micromechanics

Mario Rueda-Ruiz[1,2], Miguel Herráez[3], Federico Sket[1], Francisco Gálvez[2], Carlos González[2,1] and Jon M. Molina-Aldareguia[4,1*]

[1]IMDEA Materials Institute, C/Eric Kandel 2, 28906 Getafe, Madrid, Spain

[2]Department of Materials Science, Universidad Politécnica de Madrid, E. T. S. de Ingenieros de Caminos, 28040 Madrid, Spain

[3]TecnoDigital School

[4]Mechanical Engineering Department, Universidad Politécnica de Madrid, 28006 Madrid, Spain

**Abstract**

The use of composite materials for structures subjected to impacts requires a deep understanding about the dynamic behaviour of the material. To this end, a physically-based computational micromechanics simulation tool has been developed to predict failure initiation in a composite ply over a wide range of strain rates. The computational micromechanics framework incorporates constitutive models for the fibre, matrix and fibre-matrix interface, which are partly calibrated with novel micromechanical testing techniques. The simulation tool was applied to two matrix-dominated deformation modes of the ply: transverse compression and in-plane shear. The comparison of simulation and experimental results at coupon level have revealed a change in failure initiation mechanism of the composite ply with strain rate, which was then corroborated through observation of the fracture surfaces on the samples.

## 1. INTRODUCTION

Fibre-reinforced polymer composite materials are nowadays widely used in applications requiring high structural performance and low weight, like in the aerospace industry. However, there are areas where these materials have not reached their full potential yet, such as in components subjected to impact and high-speed events. This is mostly due to the insufficient knowledge about their dynamic behaviour and the lack of accurate analysis methodologies.

---

[*] Corresponding author: jon.molina@imdea.org; IMDEA Materials Institute, C/ Eric Kandel 2, 28906 Getafe, Madrid, Spain

There are studies in literature that have focused on the determination of failure envelopes at different strain rates, which enables the prediction of the failure load in a composite ply under combined loading conditions [1–4]. Other researchers have opted for studying the effect that strain rate has on the different failure mechanisms individually. Studies about damage on the longitudinal direction have found remarkable differences between the behaviour in tension and compression. Taniguchi et al. found that the tensile properties of carbon fibre reinforced laminates are rate insensitive [5], in line with the results of another study on carbon fibres [6]. In contrast, studies on glass fibre reinforced composites have found a pronounced rate effect [7], which is due to the sensitivity of glass fibres to the strain rate. Under longitudinal compression, the matrix and fibre-matrix interface play a more important role in failure, thus higher rate sensitivity is expected, as shown by several authors [8,9]. In the case of matrix dominated modes (transverse direction and shear modes), previous works have found significant rate dependent response [3,5,10].

Recently, a novel approach for the prediction of the mechanical behaviour in fibre-reinforced polymer composites has become popular, based on a bottom-up multiscale modelling strategy [11]. The approach is based on the transfer of information in between scales by homogenisation into constitutive models, rather than on coupling models with different levels of detail, as it was done traditionally. Physically-based models that capture the real failure mechanisms of the material are incorporated at each scale of the analysis. The first level of the analysis is the microscale or ply level. Computational micromechanics models are used to capture the level of detail of the composite microstructure, including fibres, matrix and fibre-matrix interface. One of the computational micromechanics techniques is the Representative Volume Element (RVE) approach. It is used to predict failure initiation of the composite ply in the form of failure envelopes [12–15], which are equivalent to the ones obtained by classical failure criteria of composites. An important factor that determines the accuracy of the predictions made by computational micromechanics models is the calibration of the material properties of the composite constituents. An approach that has shown particular success in previous studies is the use of micromechanical testing techniques for the *in situ* determination of matrix and fibre-matrix interface properties [16]. This way, the effects that processing conditions may have on the mechanical response of the matrix and the fibre-matrix interface are captured.

Micromechanical modelling studies on high strain rate behavior of polymer composites are still scarce. However, following the current trend towards multiscale modelling of composites, several authors have started studying the dynamic properties of composites from a multiscale modelling perspective. Lidgett et al. set up a microscale unit cell with hexagonal packing arrangement and analysed it with FEM [17]. They introduced different constitutive models for the fibre, matrix and fibre-matrix interface and calibrated the interface model with dynamic fibre push-out test results from other authors [18]. Sorini et al. used a semi-analytical micromechanics model, the Generalized Method of Cells, to carry out predictions of the ply behaviour under several loading configurations [19]. The constitutive models were calibrated with macroscale neat resin tests. Li et al. opted for the RVE approach to perform a computational micromechanics study at high strain rates [20]. They used two types of RVEs: one with a single fibre (squared arrangement) and the other with a random distribution of fibres. Again, macroscale testing data was used for model calibration. Wang et al. also developed a micromechanical model based on the RVE approach to study the transverse deformation of a composite ply from low to high strain rates [21]. They compared the behavior of random RVEs with microscale unit cell models comprising several types of fibre arrangements.

In the present study, a physically-based micromechanical modelling approach is presented that incorporates constitutive models for the fibre, matrix and fibre-matrix interface. Rate dependency is accounted for in the material model of the polymer matrix, which is partly calibrated with novel micromechanical testing techniques that cover a wide range of strain rates ($10^{-3}$ to $10^2$ s$^{-1}$). This way, this work aims at predicting damage initiation of a composite ply as a function of strain rate. Unlike previous literature works on this topic that make use of macroscale experiments for calibration of the rate dependent matrix model, this work approaches the calibration of the matrix constitutive model with novel *in-situ* micromechanical testing techniques.

## 2. MICROMECHANICAL MODELLING

### 2.1 Set-up of RVE model

The representative volume element (RVE) approach was used in the framework of the Finite Element Method (FEM) to generate a micromechanical model of a composite ply. Statistically representative microstructures were generated using a Python-based tool called VIPER [22]. This tool implements several algorithms for random microstructure generation. The algorithm developed by Herráez et al. based on overlapping potentials was used [22]. FEM models of the RVEs were created in the framework of the commercial FEM software package Abaqus. Figure 1 shows the three RVEs that were generated to account for the effects of different fibre distributions. The fibre volume fraction is 60 % and the fibre diameter 5.2 μm, as specified in the product datasheet of IM7 fibres [23], which is the reinforcement of the composite material used for validation. All the RVEs have the same size, 40x40x0.5 μm, that results in a total of 40-50 fibres. This ensures that the mechanical response of the ply is size independent, as it has been shown in previous works [24,25]. An RVE of smaller size would exhibit an influence of fibre distribution.

The domain was meshed with C3D8 and C3D6 elements, with an average element size of 0.5 μm. This means that there is only one element in the thickness direction. For modelling the fibre-matrix interface, cohesive interactions were used. They were placed in all contact surfaces between fibres and matrix, as specified in the example of Figure 1. The cohesive interactions link the degrees of freedom of the nodes that belong to the fibres with the nodes that belong to the matrix. The interaction is represented by a traction-separation law, which is defined by the material properties of the interface.

Periodic boundary conditions (PBCs) were implemented in the FEM model to establish continuity of displacements between neighbouring RVEs. The PBCs impose a relative displacement between nodes in opposite faces of the RVE equal to the relative displacement between the master nodes located at each of the surfaces. The imposed relative displacements between nodes in opposite faces are then:

$$u(L,y,z) - u(0,y,z) = U^{MN_x} - U^{MN_0} \quad \text{Equation 1.1}$$
$$u(x,L,z) - u(x,0,z) = U^{MN_y} - U^{MN_0} \quad \text{Equation 1.2}$$
$$u(x,y,W) - u(x,y,0) = U^{MN_z} - U^{MN_0} \quad \text{Equation 1.3}$$

Where L and W are the side and thickness of the RVE, respectively. The z-axis is in the longitudinal direction of the fibres. The PBCs were implemented in the FEM model automatically, using a Python code that generates equations that relate the degrees of freedom of the nodes in opposite surfaces. The nodes in opposite surfaces are located at the same position along the edge for this purpose. In addition to the PBCs, periodicity of the generated microstructure is ensured by having continuity of the fibres cut by the domain borders. Fibres sectioned by an edge have the complementary piece located on the opposite edge, as it can be seen in Figure 1.

Three load cases have been simulated: transverse compression, shear perpendicular to the fibres and shear parallel to the fibres. The load cases are introduced in the RVE by imposing displacements to the master nodes. The displacements are compiled in Table 1.

The deformation of the RVE leads to reaction forces in each of the master nodes. From the imposed displacements and the resulting reaction forces, the homogenised stress-strain response of the RVE can be determined. The in-plane shear response was obtained from the average stress of the two shear cases, perpendicular and parallel to the fibres, as it was recommended by Totry et al [26]:

$$\tau_{12} = \frac{\tau_\parallel + \tau_\perp}{2} \quad \text{Equation 2}$$

The stresses and strains in the two shear cases were computed as:

$$\gamma_\parallel = \tan^{-1}\left(\frac{U^{MN_z}}{L}\right), \quad \tau_\parallel = \frac{R^{MN_z}}{L \cdot W} \quad \text{Equation 3.1}$$

$$\gamma_\perp = \tan^{-1}\left(\frac{U^{MN_y}}{W}\right), \quad \tau_\perp = \frac{R^{MN_y}}{L \cdot L} \quad \text{Equation 3.2}$$

Stress and strain in the transverse compression case were computed as:

$$\epsilon = \frac{U^{MN_x}}{L}, \quad \sigma = \frac{R^{MN_x}}{L \cdot W} \quad \text{Equation 4}$$

where $U^{MN_x}$ and $R^{MN_x}$ are the imposed displacement and resulting reaction force, respectively, of the master node located along the x-axis. The same applies for $U^{MN_y}$ and $R^{MN_y}$, which are the imposed displacement and resulting reaction force in the master node located along the y-axis, and $U^{MN_z}$ and $R^{MN_z}$, which correspond to the master node along the z-axis.

Before the mechanical loading step, the RVEs were subjected to a thermal step that emulates the cooling process that the material undergoes after curing. A thermal step of -160 ºC was applied to the RVEs, while rigid body motions were prevented.

Regarding the applicability of micromechanical simulations for high strain rate cases, this study assumes that the RVEs and the tested samples used for validation are both in dynamic equilibrium. In other words, the test duration was sufficiently long compared to the minimum time required for a sound wave to travel one gauge length.

**2.2 Constitutive models for fibre, matrix and fibre-matrix interface**

Fibres were modelled as transversally isotropic solids, thus the mechanical behavior is defined by five elastic constants. Since the validation tests were performed on carbon-fibre composites, the mechanical properties were assumed to be rate independent, in line with the findings of previous works [6].

The mechanical behavior of the composite polymer matrix is, on the other hand, rate and pressure dependent. The constitutive model selected for the matrix takes into account these phenomena. Based on the accurate mechanical predictions of previous works on computational micromechanics of polymer composites [25,27], the matrix was modelled as an isotropic linear elastic solid with plastic behaviour defined by a Drucker-Prager yield and flow rule. The yield surface can be expressed as:

$$F(\boldsymbol{\sigma}, \beta) = q - p \tan \beta - d = 0 \qquad \text{Equation 5}$$

The parameters $\beta$ and $d$ are the frictional angle and cohesion strength, respectively. The frictional angle is related to the pressure sensitivity of the material and the cohesion is related to the yield stress. $\boldsymbol{\sigma}$ is the Cauchy stress tensor. $q$ and $p$ are the von Mises equivalent stress and hydrostatic pressure stress, respectively, which are defined by:

$$q = \sqrt{\frac{3}{2}(\boldsymbol{\sigma}':\boldsymbol{\sigma}')}, \quad q = -\frac{1}{3}tr(\boldsymbol{\sigma}) \qquad \text{Equation 6}$$

where $\boldsymbol{\sigma}'$ is the deviatoric part of the stress tensor, $\boldsymbol{\sigma}' = \boldsymbol{\sigma} + p\boldsymbol{I}$.

The cohesion strength can be expressed as a function of the compression yield stress and frictional angle of the material as $d = \sigma_{yc}(1 - 1/3 \tan \beta)$.

The Drucker-Prager model implemented in the Abaqus' material library was used. The material model requires the input of four material parameters: two elastic constants (elastic modulus, $E_m$, as a function of strain rate, and Poisson's ratio, $\nu_m$) and two parameters related to plasticity (frictional angle, $\beta$, and compression yield stress, $\sigma_{yc}$). The friction angle was assumed to be strain rate independent. The compression yield stress was assumed to be a function only of the strain rate. No plastic strain hardening was implemented, following the assumptions of previous works in literature [25,27].

The fibre-matrix interface is represented by a cohesive zone model. The separation of the fracture surfaces under the action of a traction vector is defined by a traction-separation law and damage progressively advances along the cohesive zone. Damage evolution is controlled by a scalar damage variable, *d*, that goes from 0 (no damage) to 1 (fully damaged). The model accounts for several failure modes acting simultaneously; hence, the traction stress vector ($t_n$, $t_{t1}$, $t_{t2}$) and the separation vector ($\delta_n$, $\delta_{t1}$, $\delta_{t2}$) have three components: one normal to the surface and two shear components (longitudinal and transversal). In this case, the traction-separation law can be expressed in terms of effective traction and separation:

$$\bar{t} = \sqrt{\langle t_n \rangle^2 + t_{t1}^2 + t_{t2}^2}, \quad \bar{\delta} = \sqrt{\langle \delta_n \rangle^2 + \delta_{t1}^2 + \delta_{t2}^2} \qquad \text{Equation 7}$$

The normal components use the Macaulay brackets to ensure that there is no damage advancement when the interface is in compression.

A linear traction-separation law like the one in Figure 2 was imposed for the cohesive interactions on the fibre-matrix interface. Its behaviour before damage initiates is linear and is defined by a slope *K*, which is large enough so as to preserve displacement continuity through the interface, but not too large to prevent numerical instabilities. The onset of damage is at the point where the traction-separation relation reaches its maximum. A quadratic interaction criterion was selected to determine the initiation of damage under mixed mode conditions:

$$\left(\frac{\langle t_n \rangle}{t_n^0}\right)^2 + \left(\frac{t_{t1}}{t_{t1}^0}\right)^2 + \left(\frac{t_{t2}}{t_{t2}^0}\right)^2 = 1 \qquad \text{Equation 8}$$

where $t_n^0$, $t_{t1}^0$ and $t_{t2}^0$ are the normal and the two shear strengths of the interface, respectively. Once damage initiates, the stiffness degrades as damage develops. Since a linear softening law is imposed, the damage variable can be estimated from $d = \bar{\delta}_f(\bar{\delta} - \bar{\delta}_0)/\bar{\delta}(\bar{\delta}_f - \bar{\delta}_0)$. $\bar{\delta}_f$ is the displacement jump at the point where the interface has failed completely, and it can be estimated from the effective fracture energy and strength as $2\bar{G}_c/\bar{t}_c$. A Benzeggagh-Kenane model was selected to account for mixed mode fracture [28], which is already implemented in Abaqus [29].

In this study, the cohesive zone model for simulation of interface debonding is coupled with a Coulomb friction model to reproduce the frictional interaction between the debonded surfaces after failure, following the same approach as in the work of Naya et al. [27,30].

## 3. EXPERIMENTAL PROCEDURES

### 3.1 Material

The micromechanical modelling approach was validated against experimental results of an IM7/8552 composite system. The carbon fibre HexTow IM7 of Hexcel (USA) is an intermediate modulus fibre with widespread application in industry. The epoxy resin Hexcel 8552 is an amine-cured epoxy containing polyethersulphone. Combined, the IM7/8552 is a high-performance composite material used in numerous applications, like in aerospace industry. Its properties are well known to industry; hence, it is the perfect material for validation of the micromechanical testing methodology.

Panels of this material were prepared by hand lay-up, followed by vacuum bagging and autoclave consolidation. A curing cycle including a hold at 180 ºC for 135 minutes was applied to the material, together with a pressure of 7 bar.

### 3.2 Micromechanical testing of composite constituents for model calibration

Micromechanical testing was used to determine some of the material properties of the matrix and fibre-matrix interface. This type of tests allows *in-situ* extraction of properties in the composite ply; hence, the influence that the manufacturing process has on the mechanical properties can be captured. In addition, micromechanical tests are faster and require much less material than conventional macroscale testing techniques.

The polymer matrix was tested using the micropillar compression technique to determine the compressive yield stress. The pillars were carved in the polymer surface by Focused Ion Beam (FIB) milling using a Helios NanoLab DualBeam 600i of Thermo Fisher Scientific (USA). Pillars of about 7 µm diameter were produced, which lead to size-independent properties according to previous studies [27]. They were tested in an instrumented nanoindenter NanoTest Alpha of Micro Materials Ltd (UK) equipped with a 10 µm flat punch indenter. This instrument enables mechanical characterization over a wide range of strain rates because of its pendulum-based configuration that allows conventional force-actuated nanoindentation and impact nanoindentation. A modified version of this instrument was used that incorporates a piezoelectric force sensor to measure the applied force during impact to account for the important inertia contribution [31]. The micropillars were tested at 10 and 100 nm/s in the low-speed configuration, and 0.5 and 0.9 mm/s in the impact configuration. More detailed descriptions of the test and results can be found in previous works [32].

The fibre push-in test was used for determination of the shear strength of the fibre-matrix interface. The nanoindentation instrument was equipped with a 5 µm flat punch indenter, which is just 0.2 µm smaller than the fibre diameter. This eliminated the risk of plastically deforming the fibre during indentation. A cross-section of a composite laminate was prepared in such a way that there were several plies with fibres perpendicular to the surface. The surface was ground and polished to a fine degree. Tests were performed in an instrumented nanoindenter Hysitron TI-950 TriboIndenter of Bruker (USA). A displacement-controlled scheme at 30 nm/s was applied. The high-speed tests using the impact configuration of the NanoTest Alpha instrument could not be used in this case because of limitation of the test set-up, as described in detail in [33]. Around ten push-in tests were performed for statistical purposes. The results were analysed using a FEM-based methodology developed by Rodríguez et al. [34] and Naya [27].

**3.3 Macroscale testing for RVE model validation**

The micromechanical modelling methodology was validated against macroscale experimental results at the coupon (ply) level. Two loading configurations were tested, transverse compression and in-plane shear, as in the simulations. These were selected because the damage mechanisms are controlled by the matrix and fibre-matrix interface, which exhibit a rate sensitive mechanical response. Therefore, the composite ply was expected to show rate sensitivity too.

A test campaign was prepared to determine the mechanical response of the ply over a wide range of strain rates (from $10^{-3}$ to $10^3$ s$^{-1}$). This required the use of several mechanical testing frames because one single device cannot cover such wide strain rate range. Low strain rate testing (up to 1-10 mm/s) was carried out in a screw-driven electromechanical frame Instron 5966 equipped with a 10 kN load cell. Mechanical wedge grips were used for the in-plane shear tests. Low and intermediate strain rates (up to 1000 mm/s) were performed in a servo-hydraulic machine Gleeble 3800 of Dynamic Systems (USA) equipped with a 250 kN load cell. Custom-built mechanical wedge grips were used for the in-plane shear tests at intermediate rates (10-1000 mm/s). Transverse compression tests at low and intermediate strain rates were carried out with two self-aligning plates.

The highest strain rate regime (around $10^4$ mm/s) was covered with a split-Hopkinson pressure bar (SHPB). Two types of bars were used. A compression bar for the transverse compression tests and a tensile SHPB set-up for the in-plane shear. Load was transferred by contact in the transverse compression specimens. For the in-plane shear samples, custom-built grips were made with a notch where the tabbed area of the sample was inserted and glued (Loctite EA 9466). All bars of the SHPB set-ups were made of Inconel and both the incident and transmitted bars were instrumented with strain gauges in Wheatstone bridge configuration to measure the strain pulses.

Videoextensometry was used to analyse the fracture mechanisms and perform digital image correlation (DIC) measurements. A Stingray F-504 camera of Allied Vision (Germany) equipped with Pentax C7528-M 75 mm f2.8 optics was used for the low speed tests. Intermediate rate testing in the hydraulic frame was accompanied with a Phantom VEO 1310 of Vision Research (USA) equipped with AF Zoom-Nikkor 24-85 mm f/2.8-4D IF optics. A Phantom VEO 710 of Vision Research (USA) equipped with Micro-Nikkor 105 mm f/2.8 optics was used in the SHPB set-up.

Transverse compression samples consisted of end-loading 20x10x5 mm prisms. They were extracted from 5 mm thick panels with $[90]_{20}$ lay-up by CNC machining. Tests were performed at 0.2, 20, 1000 and ~$1.5 \cdot 10^4$ mm/s. The in-plane shear response was studied by tensile testing of ±45 samples. In addition, the samples were dogbone shaped, with dimensions optimized for SHPB testing. Gauge length is 10x5 mm, and the total length and width of the sample is 73 and 20 mm, respectively. Specimens were extracted from 2 mm thick laminates ($[\pm 45]_{2s}$). Tests were performed at 0.02, 2, 200 and ~$2 \cdot 10^4$ mm/s. Shear stress-strain curves were obtained following the data analysis procedures in the ASTM D3518 standard: $\gamma_{12} = \epsilon_x - \epsilon_y$, $\tau_{12} = P/2A_0$. $\epsilon_x$ and $\epsilon_y$ are the strain along the loading axis and perpendicular to the loading axis, respectively, and were extracted from the DIC measurements.

## 4. MICROMECHANICAL SIMULATIONS AND CORRELATION WITH EXPERIMENTS

### 4.1 Summary of inputs of the material models

The elastic constants of the fibre were extracted from an inverse analysis of macroscale testing results of composite and neat resin coupons. Chamis rule of mixtures was employed [35], following the recommendations of previous studies on computational micromechanics of composites [25,27]. The thermal expansion coefficients were taken from the work of Kaddour and Hinton [36]. The values are summarized in Table 2.

The material properties used for the matrix material model are included in Table 3. The elastic modulus was defined as a power law function of the strain rate, which was extracted from Dynamic Mechanical Analysis (DMA) results scaled by the elastic modulus obtained in macroscale tensile tests of 8552 in neat resin form [33]. The Poisson's ratio was also obtained from the same macroscale tensile tests. The compression yield stress was obtained from micropillar compression tests in neat resin coupons, as described in section 3.2. More details can be found elsewhere [32]. A bilinear law was derived, which is typical of epoxy resins. The frictional angle and thermal expansion coefficient were obtained from literature [25,27].

The material properties of the fibre-matrix interface are compiled in Table 4. The longitudinal shear strength was obtained *in-situ* from the push-in tests described in section 3.2. The transversal shear strength was assigned the same value, and a 2/3 factor was applied to estimate the normal interface strength, following recommendations of Naya [27]. The interface fracture energies were also taken from the work of Naya [27]. The friction coefficient was adjusted to get an optimum fit of the transverse compression results, which are more sensitive to this parameter than the in-plane shear loading.

The simulation of different strain rates was carried out by adjusting the material properties for each strain rate case, instead of running simulations with different step times and a rate dependent material. Introducing rate dependency in the FEM models led to convergence issues. The validity of this approach was checked by comparing results of models run with several step times and rate dependent material models, with models run at a constant step time and rate independent material models in which the properties were adjusted in each case of strain rate. Both approaches delivered the same results, but convergence was better in the analysis with no rate dependent material models. The Abaqus Standard (Implicit) solver was used. Better convergence could be potentially achieved with the Explicit solver. However, modelling of the lowest strain rate cases with the Explicit solver would require a high degree of mass scaling to reach a feasible analysis time.

The compression yield stress and elastic modulus were the only inputs changed in the material model of the matrix. They were those gathered in Table 3, by applying the strain rate that corresponds to each simulation case.

**4.2 In-plane shear behaviour**

Figure 3 presents the outcomes of the FEM simulations using the material set of the IM7-8552 composite described in section 4.1. The results of FEM simulations using a perfect interface (no damage) have been also included. The simulation results were compared with experimental results from in-plane shear tests covering a wide range of strain rates, which are described in section 3.3. The mechanical response of the ply predicted by the FEM simulation correlates well with the experimental result for the lower speed case. The simulation using a perfect interface shows that the ply could still carry higher loads if the interface had not failed. This means that damage initiation under in-plane shear loading for the lowest strain rate case is dominated by fibre-matrix interface debonding. The other two higher strain rate cases show a significant deviation between the experimental trends and the simulation results using the interface properties in Table 4. Furthermore, the gap increases for higher strain rates, suggesting that the strain rate dependent mechanical behaviour of the interface is not properly represented by the material properties in Table 4 that describe a rate-insensitive interface.

Interestingly, the simulation results using a perfect interface model reproduce accurately the experimental trends up to a certain strain point in the two higher strain rate cases. This suggests that not only the mechanical behaviour of the interface is rate dependent, but that the shear strength of the interface actually increases at a higher rate with strain rate than the shear strength of the matrix. This then leads to a change in the damage initiation mechanism from damage initiated by combined interface debonding and matrix plasticity at the lowest strain rate case to damage initiated by matrix plasticity exclusively for the other two higher strain rate cases.

In addition to the predicted homogenised stress-strain response of the ply, the FEM simulations allow a more in-depth analysis of the damage initiation mechanisms by looking at the distributions of stress and strain in the RVE. Figure 4 shows contour plots of the equivalent plastic strain at the point of maximum simulated deformation for the two more extreme strain rate cases. The RVEs of the high strain rate case correspond to simulations using a perfect interface because this seems to be the representative behaviour of the ply at this strain rate, according to the results of Figure 3. In the case of shear parallel to fibres at low strain rates, the matrix exhibits plastic deformation along the entire RVE, but there is a dominant band of higher plastic deformation parallel to the shear loading. This area of localized plastic strain was induced by the debonding of the fibre-matrix interface. When the interface fully fails, the fibre cannot carry load anymore, thus increasing the stress in the surrounding matrix. On the other hand, at high strain rates there is no band of localized plastic deformation and the plastic strains are uniformly distributed along the microstructure. This is because there is no interface debonding that may trigger localization of damage. As expected, the plastic strains in the high strain rate case are higher than at low strain rate because the stress carried by the ply is higher. The case of shear perpendicular to fibres exhibits a similar response than the shear parallel to fibres; although, the strain localization at low strain rate is not as obvious under this loading condition. Plastic deformation is generally more widespread along the microstructure in the case of shear perpendicular to the fibres [37]. At high strain rates, the response is nearly the same as for shear parallel to fibres.

**4.3 Characterization of the fracture surface**

The fracture surfaces of the in-plane shear samples were analysed in a scanning electron microscope (Helios NanoLab DualBeam 600i of Thermo Fisher Scientific, USA) to have an insight about the dominant fracture mechanisms in the samples and the effect of strain rate. Figure 5 compiles representative images of the fracture morphology for each strain rate case. At $0.002 \text{ s}^{-1}$, there is a clear predominance of fibre-matrix interface debonding materialized in the imprints left by the fibres in the epoxy matrix. The morphologies of the intermediate and high strain rate cases, 0.2 and $1000 \text{ s}^{-1}$ respectively, are much alike among each other, but show significant differences with the fracture surfaces of the low speed case. There is much less amount of fibre-matrix debonding. The fracture surface is dominated by the highly deformed matrix, which is materialized in a much rougher fracture surface with less imprints left by the fibres.

These observations are thus in line with the conclusions obtained by comparing the simulations results with the macroscale experiments. There is a change in the fracture initiation mechanism with strain rate that goes from fracture dominated by combined fibre-matrix debonding and matrix plasticity at low strain rates, to fracture dominated mostly by matrix plasticity at high strain rates. These findings are also in agreement with the observations made by other authors regarding the change in damage initiation mechanism under in-plane shear over a wide range of strain rates. By looking at the fracture surfaces, both Cui et al. and Taniguchi et al. concluded that cracks propagated primarily along the matrix at high strain rates; while at low strain rate the cracks propagated mainly along the fibre-matrix interface [5,10].

**4.4 Transverse compression behaviour**

Figure 6 presents the results of the FEM simulations, which are compared with experimental results from the test campaign covering a wide range of strain rates (see section 3.3). The results of FEM simulations considering a perfect interface have been also included. At low strain rate, the simulations featuring an interface with the properties in Table 4 exhibit a close match with the experimental trends, meaning that at low strain rate damage initiation is controlled by combined matrix plasticity and fibre-matrix debonding. Nevertheless, the stress-strain response of the ply is just a bit deviated from the perfect interface case, so mainly plastic deformation dominates the nonlinear deformation. The high strain rate results show the opposite behaviour. The prediction of the perfect interface model matches better the experimental trend in this case, which is in line with the in-plane shear response described before. Therefore, these results provide additional evidence to support the idea that there is a change in the damage initiation mechanism from combined matrix plasticity and interface debonding at low strain rates to just matrix plasticity at high strain rates.

The contour plots in Figure 7 give deeper insight on the damage mechanisms of the ply under transverse compression. At low strain rates, the microstructure exhibits several bands of localized plastic deformation. Some of these bands form an angle with the perpendicular to the loading direction of about 55º, which is a reasonable orientation of the fracture angle for an epoxy resin [24]. This angle is related to the pressure sensitivity of the polymer matrix. There are also other bands in the microstructure that form a lower angle. This is because the fibre-matrix interface debonding triggers the plastic deformation of the matrix. Then, the bands initiated by the interface debonding all coalesce in a shear band. The contour plot of Figure 7 for the high strain rate case corresponds to the simulations with a perfect interface because this is the case that best represents the experimental trends. The contour plot shows that the deformation is nearly linear elastic for the high strain rate case. Only a few areas of high stress concentration in matrix ligaments in between fibres exhibit plastic deformation (marked with red arrows in Figure 7). High stress concentrations in the area may be caused by volumetric locking of the wedge (C3D6) elements located in the ligaments between fibers.

In line with the results of Figure 7, the fracture angle of the tested samples also form an angle of about 55º with the perpendicular to the loading direction, as it is shown in Figure 8. Interestingly, the fracture angle does not seem to change with strain rate, an observation which is in line with the findings of other researchers [38,39]. This would confirm the assumption made in this work that the pressure sensitivity parameter of the resin is not rate dependent.

## 5. CONCLUSIONS

A novel methodology has been proposed to predict the effect of strain rate on the in-plane shear and transverse compression response of a composite ply, focusing on the damage initiation stage. These are two deformation modes of the ply which are dominated by the properties of the matrix and the fibre-matrix interface, thus significant rate dependence is expected given the polymeric nature of the matrix. The proposed methodology is built around physically-based computational micromechanics simulations of the composite ply that use the Representative Volume Element approach in the framework of Finite Element Modelling. One of the unique features of this methodology is that the material models of the composite constituents have been partly calibrated with novel micromechanical testing techniques that cover a wide range of strain rates.

The comparison of simulations and experiments at the coupon level have shown that damage initiation mechanisms change with strain rate for in-plane shear and transverse compression loading. At low strain rates, ply failure is controlled by the combined effect of fibre-matrix interface debonding and matrix plasticity, while at intermediate and high strain rates failure is dominated mostly by matrix plasticity. This suggests that the fibre-matrix interface is strongly rate dependent and that its rate dependency is more pronounced that the rate sensitivity of the matrix alone. Further evidence of this transition in damage initiation mechanism with strain rate was provided by the observation of the fracture surfaces in the in-plane shear tests at the three tested strain rates. There is a clear change in fracture surface morphology when comparing the low strain rate case with the other two higher strain rate cases.

This work was aimed primarily at studying the role of strain rate on damage initiation of composite plies. Future works should aim at predicting the final failure (and strength), for which the implementation of a strain rate depending damage model for the matrix would be required.

**Acknowledgements**

This paper is dedicated to the memory of our dear colleague Cláudio S. Lopes, who recently passed away. He made a great impact at IMDEA Materials Institute and he is deeply missed.

The research leading to these results received funding from the European Union's Horizon 2020 research and innovation programme under the Marie Sklodowska-Curie grant agreement Nº 722096, DYNACOMP project.

**Data availability statement**

The datasets generated during and/or analysed during the current study are available from the corresponding author on reasonable request.

**Competing interests statement**

The authors declare that they have no competing interests.

**REFERENCES**

[1] Daniel IM, Werner BT, Fenner JS. Strain-rate-dependent failure criteria for composites. Composites Science and Technology 2011;71:357–64. https://doi.org/10.1016/j.compscitech.2010.11.028.

[2] Schaefer JD, Werner BT, Daniel IM. Strain-Rate-Dependent Failure of a Toughened Matrix Composite. Exp Mech 2014;54:1111–20. https://doi.org/10.1007/s11340-014-9876-0.

[3] Koerber H, Xavier J, Camanho PP. High strain rate characterisation of unidirectional carbon-epoxy IM7-8552 in transverse compression and in-plane shear using digital image correlation. Mechanics of Materials 2010;42:1004–19. https://doi.org/10.1016/j.mechmat.2010.09.003.

[4] Kuhn P, Ploeckl M, Koerber and H. Experimental investigation of the failure envelope of unidirectional carbon-epoxy composite under high strain rate transverse and off-axis tensile loading. EPJ Web of Conferences 2015;94:01040. https://doi.org/10.1051/epjconf/20159401040.

[5] Taniguchi N, Nishiwaki T, Kawada H. Tensile strength of unidirectional CFRP laminate under high strain rate. Advanced Composite Materials 2007;16:167–80. https://doi.org/10.1163/156855107780918937.

[6] Zhou Y, Wang Y, Jeelani S, Xia Y. Experimental Study on Tensile Behavior of Carbon Fiber and Carbon Fiber Reinforced Aluminum at Different Strain Rate. Appl Compos Mater 2007;14:17–31. https://doi.org/10.1007/s10443-006-9028-5.

[7] Gerlach R, Siviour CR, Wiegand J, Petrinic N. The Strain Rate Dependent Material Behavior of S-GFRP Extracted from GLARE. Mechanics of Advanced Materials and Structures 2013;20:505–14. https://doi.org/10.1080/15376494.2011.627646.

[8] Koerber H, Camanho PP. High strain rate characterisation of unidirectional carbon–epoxy IM7-8552 in longitudinal compression. Composites Part A: Applied Science and Manufacturing 2011;42:462–70. https://doi.org/10.1016/j.compositesa.2011.01.002.

[9] Ploeckl M, Kuhn P, Grosser J, Wolfahrt M, Koerber H. A dynamic test methodology for analyzing the strain-rate effect on the longitudinal compressive behavior of fiber-reinforced composites. Composite Structures 2017;180:429–38. https://doi.org/10.1016/j.compstruct.2017.08.048.

[10] Cui H, Thomson D, Pellegrino A, Wiegand J, Petrinic N. Effect of strain rate and fibre rotation on the in-plane shear response of ±45° laminates in tension and compression tests. Composites Science and Technology 2016;135:106–15. https://doi.org/10.1016/j.compscitech.2016.09.016.

[11] LLorca J, González C, Molina-Aldareguía JM, Segurado J, Seltzer R, Sket F, et al. Multiscale Modeling of Composite Materials: a Roadmap Towards Virtual Testing. Advanced Materials 2011;23:5130–47. https://doi.org/10.1002/adma.201101683.

[12] Totry E, González C, LLorca J. Failure locus of fiber-reinforced composites under transverse compression and out-of-plane shear. Composites Science and Technology 2008;68:829–39. https://doi.org/10.1016/j.compscitech.2007.08.023.

[13]  Totry E, González C, LLorca J. Prediction of the failure locus of C/PEEK composites under transverse compression and longitudinal shear through computational micromechanics. Composites Science and Technology 2008;68:3128–36. https://doi.org/10.1016/j.compscitech.2008.07.011.

[14]  Melro AR, Camanho PP, Andrade Pires FM, Pinho ST. Micromechanical analysis of polymer composites reinforced by unidirectional fibres: Part I – Constitutive modelling. International Journal of Solids and Structures 2013;50:1897–905. https://doi.org/10.1016/j.ijsolstr.2013.02.009.

[15]  Melro AR, Camanho PP, Andrade Pires FM, Pinho ST. Micromechanical analysis of polymer composites reinforced by unidirectional fibres: Part II – Micromechanical analyses. International Journal of Solids and Structures 2013;50:1906–15. https://doi.org/10.1016/j.ijsolstr.2013.02.007.

[16]  Herráez M, Naya F, González C, Monclús M, Molina J, Lopes CS, et al. Microscale Characterization Techniques of Fibre-Reinforced Polymers. In: Beaumont PWR, Soutis C, Hodzic A, editors. The Structural Integrity of Carbon Fiber Composites: Fifty Years of Progress and Achievement of the Science, Development, and Applications, Cham: Springer International Publishing; 2017, p. 283–99. https://doi.org/10.1007/978-3-319-46120-5_10.

[17]  Lidgett M, Brooks R, Warrior N, Brown KA. Virtual modelling of microscopic damage in polymer composite materials at high rates of strain. Plastics, Rubber and Composites 2011;40:324–32. https://doi.org/10.1179/1743289810Y.0000000007.

[18]  Tanoglu M, McKnight SH, Palmese GR, Gillespie JW. A new technique to characterize the fiber/matrix interphase properties under high strain rates. Composites Part A: Applied Science and Manufacturing 2000;31:1127–38. https://doi.org/10.1016/S1359-835X(00)00070-1.

[19]  Sorini C, Chattopadhyay A, Goldberg RK. An improved plastically dilatant unified viscoplastic constitutive formulation for multiscale analysis of polymer matrix composites under high strain rate loading. Composites Part B: Engineering 2020;184:107669. https://doi.org/10.1016/j.compositesb.2019.107669.

[20]  Li Z, Ghosh S. Micromechanics modeling and validation of thermal-mechanical damage in DER353 epoxy/borosilicate glass composite subject to high strain rate deformation. International Journal of Impact Engineering 2020;136:103414. https://doi.org/10.1016/j.ijimpeng.2019.103414.

[21]  Wang M, Zhang P, Fei Q. Transverse Properties Prediction of Polymer Composites at High Strain Rates Based on Unit Cell Model. J Aerosp Eng 2018;31:04017102. https://doi.org/10.1061/(ASCE)AS.1943-5525.0000813.

[22]  Herráez M, Segurado J, González C, Lopes CS. A microstructures generation tool for virtual ply property screening of hybrid composites with high volume fractions of non-circular fibers – VIPER. Composites Part A: Applied Science and Manufacturing 2020;129:105691. https://doi.org/10.1016/j.compositesa.2019.105691.

[23]  Hexcel. HexTow IM7 Carbon Fiber Product Data Sheet. n.d.

[24]  González C, LLorca J. Mechanical behavior of unidirectional fiber-reinforced polymers under transverse compression: Microscopic mechanisms and modeling. Composites Science and Technology 2007;67:2795–806. https://doi.org/10.1016/j.compscitech.2007.02.001.

[25]  Rodríguez M. Micromechanical characterization and simulation of the effect of environmental aging in structural composite materials. Universidad Politécnica de Madrid, 2012.

[26]   Totry E, Molina-Aldareguía JM, González C, LLorca J. Effect of fiber, matrix and interface properties on the in-plane shear deformation of carbon-fiber reinforced composites. Composites Science and Technology 2010;70:970–80. https://doi.org/10.1016/j.compscitech.2010.02.014.

[27]   Naya Montáns F. Prediction of mechanical properties of unidirectional FRP plies at different environmental conditions by means of computational micromechanics. phd. E.T.S.I. Caminos, Canales y Puertos (UPM), 2017.

[28]   Benzeggagh ML, Kenane M. Measurement of mixed-mode delamination fracture toughness of unidirectional glass/epoxy composites with mixed-mode bending apparatus. Composites Science and Technology 1996;56:439–49. https://doi.org/10.1016/0266-3538(96)00005-X.

[29]   Simulia. Abaqus Theory Manual 2016:1172.

[30]   Naya F, González C, Lopes CS, Van der Veen S, Pons F. Computational micromechanics of the transverse and shear behavior of unidirectional fiber reinforced polymers including environmental effects. Composites Part A: Applied Science and Manufacturing 2017;92:146–57. https://doi.org/10.1016/j.compositesa.2016.06.018.

[31]   Rueda-Ruiz M, Beake BD, Molina-Aldareguia JM. New instrumentation and analysis methodology for nano-impact testing. Materials & Design 2020;192:108715. https://doi.org/10.1016/j.matdes.2020.108715.

[32]   Rueda-Ruiz M, Monclús MA, Beake BD, Gálvez F, Molina-Aldareguia JM. High strain rate compression of epoxy micropillars. Extreme Mechanics Letters 2020;40:100905. https://doi.org/10.1016/j.eml.2020.100905.

[33]   Rueda Ruiz M. Experimental and computational micromechanics of fibre-reinforced polymer composites at high strain rates. phd. E.T.S.I. Caminos, Canales y Puertos (UPM), 2021.

[34]   Rodríguez M, Molina-Aldareguía JM, González C, LLorca J. A methodology to measure the interface shear strength by means of the fiber push-in test. Composites Science and Technology 2012;72:1924–32. https://doi.org/10.1016/j.compscitech.2012.08.011.

[35]   Chamis CC. Mechanics of Composite Materials: Past, Present and Future. NASA: 1984.

[36]   Kaddour A, Hinton M. Input data for test cases used in benchmarking triaxial failure theories of composites. Journal of Composite Materials 2012;46:2295–312. https://doi.org/10.1177/0021998312449886.

[37]   Totry E, González C, LLorca J, Molina-Aldareguía JM. Mechanisms of shear deformation in fiber-reinforced polymers: experiments and simulations. Int J Fract 2009;158:197–209. https://doi.org/10.1007/s10704-009-9353-4.

[38]   Vural M, Ravichandran G. Transverse Failure in Thick S2-Glass/ Epoxy Fiber-Reinforced Composites. Journal of Composite Materials 2004;38:609–23. https://doi.org/10.1177/0021998304042400.

[39]   Wiegand J. Constitutive modelling of composite materials under impact loading. http://purl.org/dc/dcmitype/Text. Oxford University, UK, 2009.

| Load case | MN$_X$ | MN$_Y$ | MN$_Z$ |
|---|---|---|---|
| Shear parallel to fibres ($\tau_\parallel$) | (0, 0, -) | (0, 0, 2.002) | (0, 0, 0) |
| Shear perpendicular to fibres ($\tau_\perp$) | (-, 0, 0) | (0, 0, 0) | (0, 0.025, 0) |
| Transverse compression ($\sigma$) | (-1.4, -, -) | (0, -, -) | (0, 0, -) |

**Table 1 – Displacements of the master nodes in each load case.**

| IM7 fibre material properties | | | | | | |
|---|---|---|---|---|---|---|
| $E_1^f$ [GPa] | $\nu_{12}^f$ | $E_2^f$ [GPa] | $G_{12}^f$ [GPa] | $G_{23}^f$ [GPa] | $\alpha_1^f$ [$K^{-1}$] | $\alpha_2^f$ [$K^{-1}$] |
| 275.7 | 0.23 | 12.2 | 18.3 | 4.7 | -4·10$^{-7}$ | 5.6·10$^{-6}$ |

**Table 2 – Inputs for the material model of the fibres.**

| 8552 matrix material properties | |
|---|---|
| Elastic modulus, $E_m$ [MPa] | $4585\dot{\epsilon}^{0.012}$ |
| Poisson's ratio, $\nu_m$ [−] | 0.3738 |
| Frictional angle, $\beta$ [°] | 30 |
| Compression yield stress, $\sigma_{yc}$ [MPa] | $\sigma_{yc}(\dot{\epsilon}) = \begin{cases} 191\dot{\epsilon}^{0.0298}, & \dot{\epsilon} < 1 \\ 189\dot{\epsilon}^{0.0536}, & \dot{\epsilon} \geq 1 \end{cases}$ |
| Coefficient of thermal expansion, $\alpha^m$ [$K^{-1}$] | 5.2·10$^{-5}$ |

**Table 3 – Inputs for the material model of the matrix.**

| IM7/8552 interface material properties | |
|---|---|
| Penalty stiffness, $K$ [$GPa/\mu m$] | 100 |
| Normal strength, $t_n^0$ [MPa] | 49 |
| Longitudinal shear strength, $t_{t1}^0$ [MPa] | 74 |
| Transversal shear strength, $t_{t2}^0$ [MPa] | 74 |
| Mode I fracture energy, $G_n$ [$nJ/\mu m^2$] | 0.002 |
| Mode II fracture energy, $G_{t1}$ [$nJ/\mu m^2$] | 0.1 |
| Mode III fracture energy, $G_{t2}$ [$nJ/\mu m^2$] | 0.1 |
| Benzeggagh-Kenane exponent | 1.45 |

| Friction coefficient | 0.1 |
|---|---|

**Table 4 – Inputs for the material model of the fibre-matrix interface.**

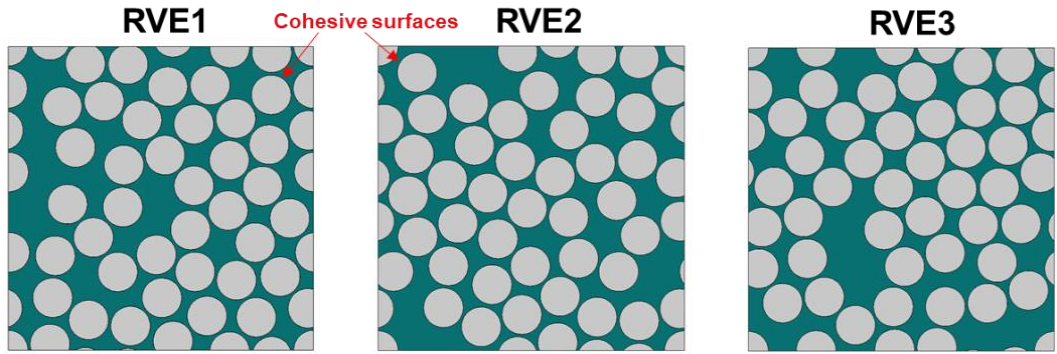

**Figure 1 - RVEs generated for the micromechanical simulations.**

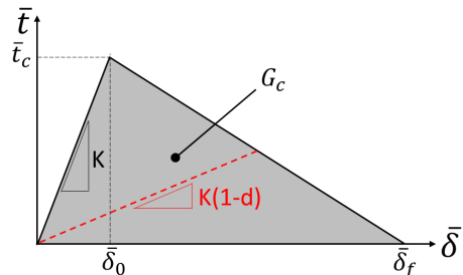

**Figure 2 – Traction separation law that defines the behavior of the cohesive zone.**

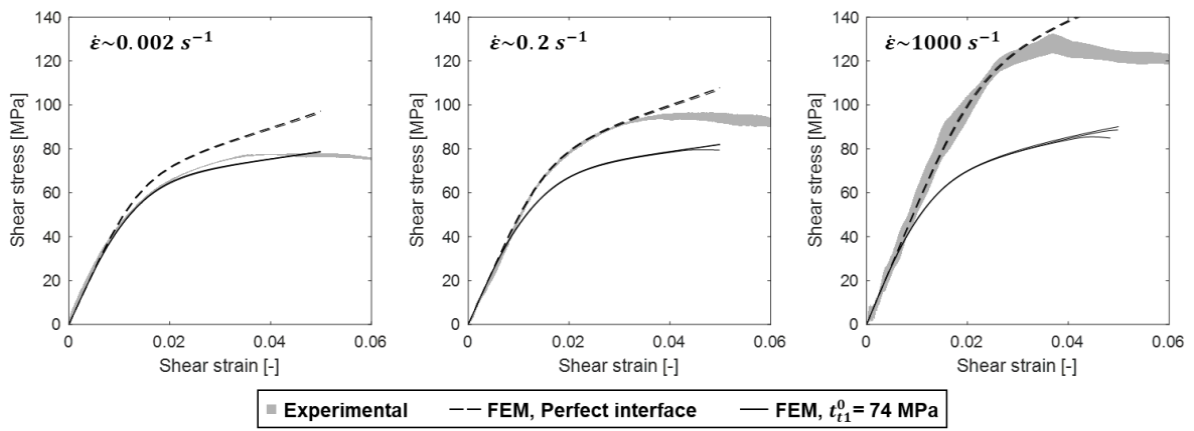

**Figure 3 – Comparison of experimental and numerical response of the in-plane shear test. Simulation results for a perfect interface are also included.**

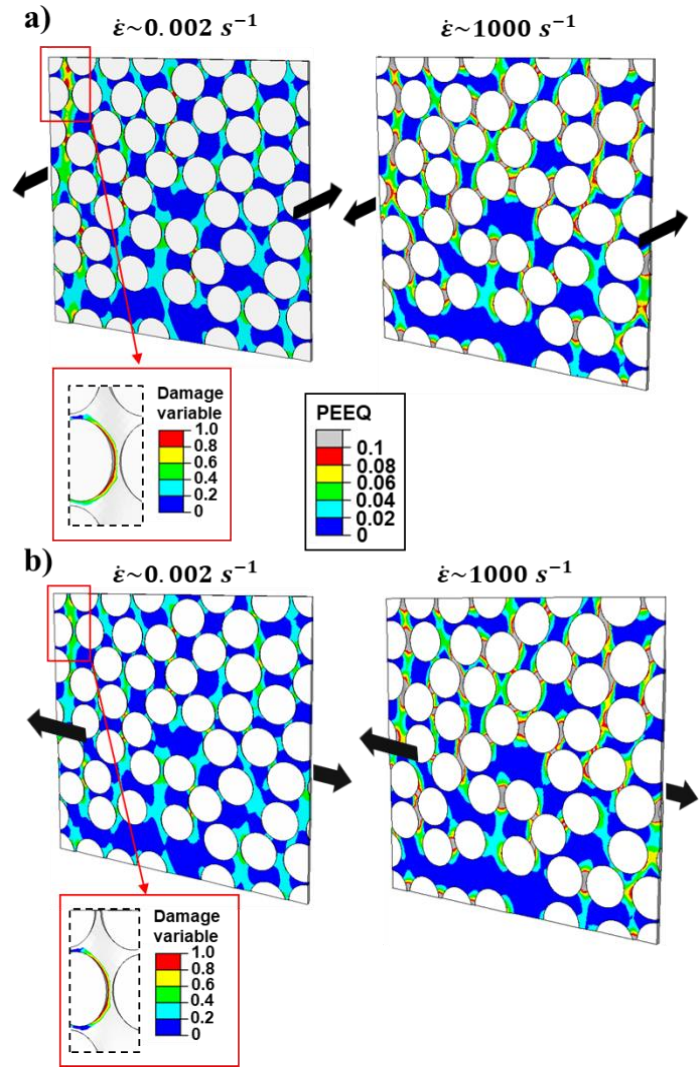

**Figure 4 - Equivalent plastic strain at maximum stress in the cases of (a) shear parallel to fibres and (b) shear perpendicular to fibres. The high strain rate cases correspond to the FEM simulations with perfect interface.**

Tests at 0.002 s$^{-1}$

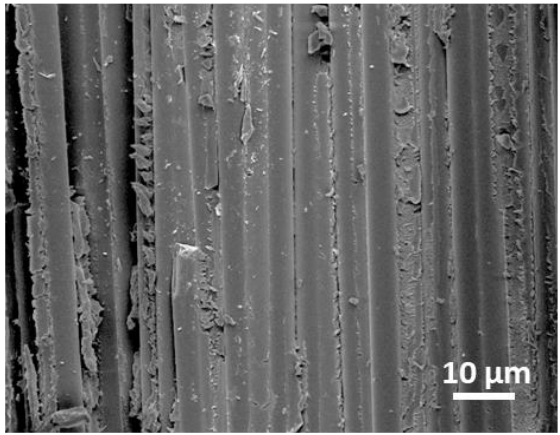
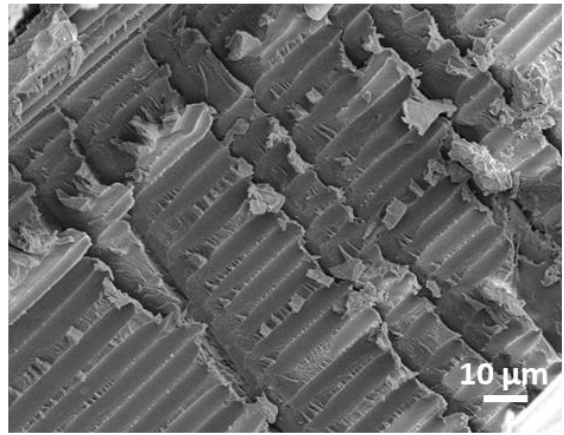

Tests at 0.2 s$^{-1}$

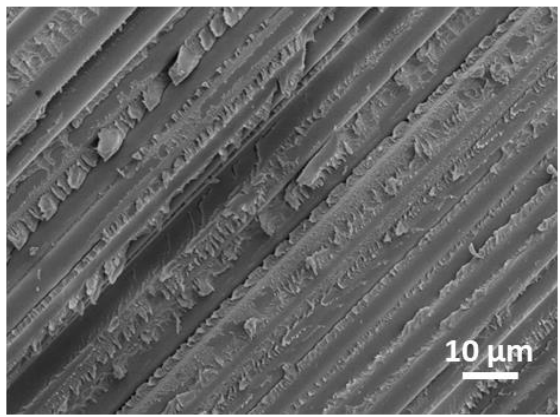
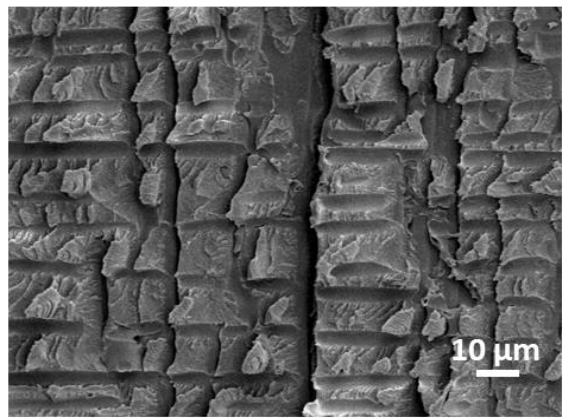

Tests at 1000 s$^{-1}$

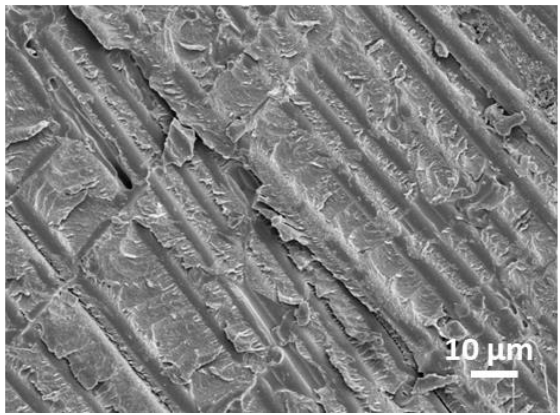
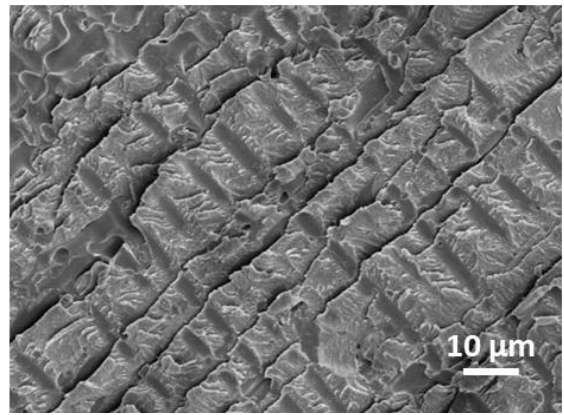

**Figure 5 – Typical fracture surfaces of the in-plane shear tests at 0.002, 0.2 and 1000 s$^{-1}$.**

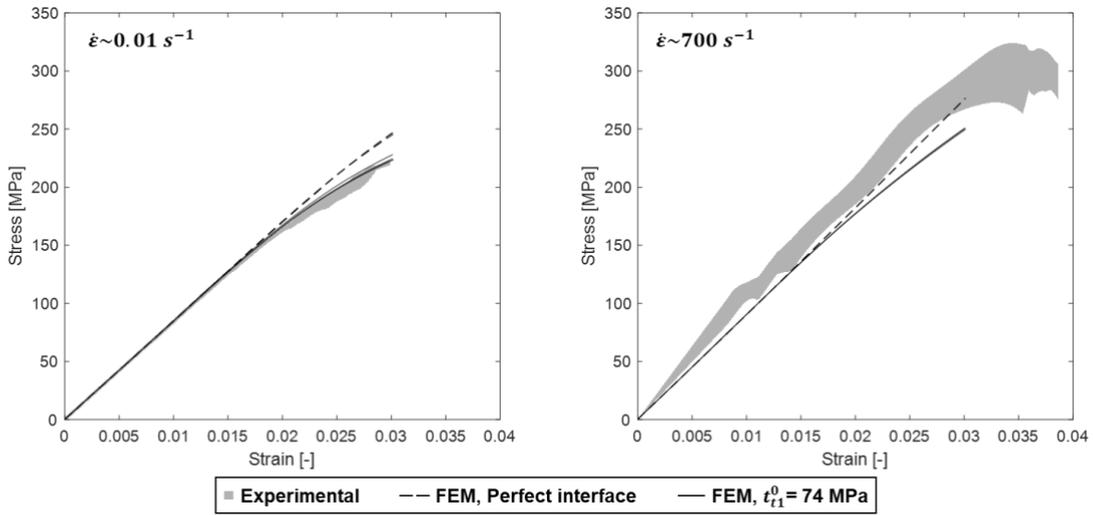

**Figure 6 -** Comparison of experimental and simulation response of the transverse compression test. Simulation results for a perfect interface are also included.

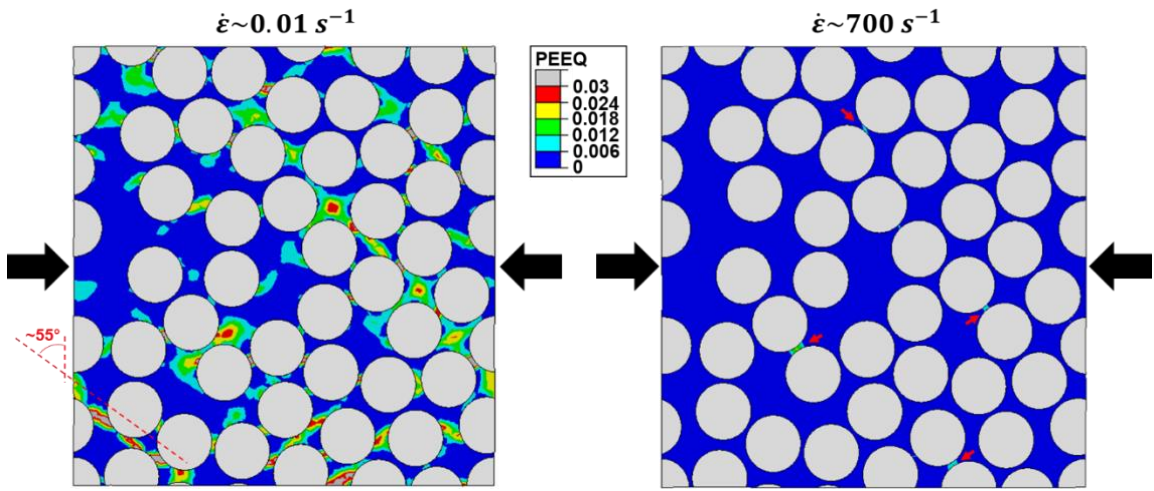

**Figure 7 -** Equivalent plastic strain at maximum stress in the case of transverse compression. The high strain rate case corresponds to the FEM simulations with perfect interface.

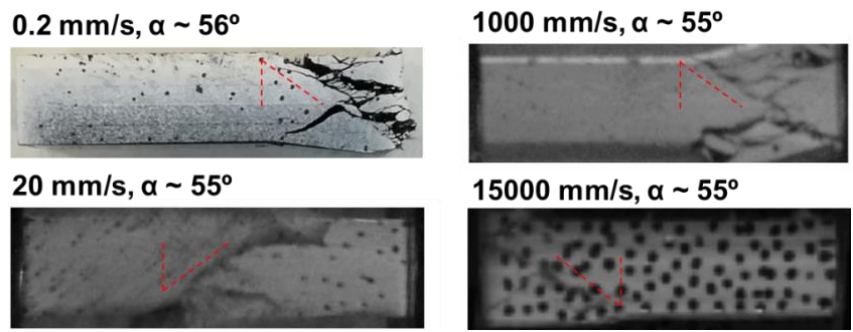

**Figure 8 -** Fracture angle in the transverse compression experiments